# DMP_AI: An AI-Aided K-12 System for Teaching and Learning in Diverse Schools

Zhen-Qun Yang[1], Jiannong Cao[1], Xiaoyin Li[1], Kaile Wang[1], Xinzhe Zheng[1], Kai Cheung Franky Poon[2] and Daniel Lai[1]

[1] Department of Computing, The Hong Kong Polytechnic University, Hong Kong
[2] HKRSS Tai Po Secondary School, Hong Kong
{zq-cs.yang, jiannong.cao, xiaoyin.li, kaile.wang, xinzhe.zheng}@polyu.edu.hk, kcpoon@hkrsstpss.edu.hk, daniel.sc.lai@hkjc.org.hk

**Abstract.** The use of Artificial Intelligence (AI) has gained momentum in education. However, the use of AI in K-12 education is still in its nascent stages, and further research and development is needed to realize its potential. Moreover, the creation of a comprehensive and cohesive system that effectively harnesses AI to support teaching and learning across a diverse range of primary and secondary schools presents substantial challenges that need to be addressed. To fill these gaps, especially in countries like China, we designed and implemented the **DMP_ AI** (Data Management Platform_Artificial Intelligence) system, an innovative AI- aided educational system specifically designed for K-12 education. The system utilizes data mining, natural language processing, and machine learning, along with learning analytics, to offer a wide range of features, including student academic performance and behavior prediction, early warning system, analytics of Individualized Education Plan, talented students' prediction and identification, and cross-school personalized electives recommendation. The development of this system has been meticulously carried out while prioritizing user privacy and addressing the challenges posed by data heterogeneity. We successfully implemented the DMP_AI system in real-world primary and secondary schools, allowing us to gain valuable insights into the potential and challenges of integrating AI into K-12 education in the real world. This system will serve as a valuable resource for supporting educators in providing effective and inclusive K-12 education.

## 1 Introduction

The elementary and secondary school stages (K-12) [1] are crucial for student development. This period is a critical phase for children's learning and growth, with profound implications for their future development. Thus, effective analysis, identification, and timely provision of support and resources to ensure students' holistic development during these formative years are essential. Artificial Intelligence (AI) has been increasingly integrated into various sectors, such as government, business, industry, education, and healthcare, revolutionizing traditional practices and introducing innovative solutions. The advent of AI technologies has opened up new possibilities for personalized learning, intelligent tutoring, and data-driven management in the educational sector.

However, despite the success of AI applications in other fields, its impact in the realm of education [2-4], remains limited. Especially for the application of AI in K-12 education is a relatively nascent field that requires further exploration and research [5,6]. The potential of AI as a tool for enhancing learning methodologies in K-12 education is vast, ranging from adaptive learning systems that cater to individual student's learning pace and style to intelligent systems that can predict student performance and identify potential areas of improvement. This requires researchers to further explore and investigate the applications of AI in current K-12 education, discover and address the challenges faced, and unlock the full potential of AI in transforming the educational landscape.

To fill these gaps, especially in countries like China, in this work, we designed and implemented the DMP_AI system, an innovative AI-aided educational system specifically designed for K-12 education in the real world to meet the increasingly diverse needs of students. It was initiated in 2021 and founded by the Hong Kong Jockey Club Charities Trust. It utilizes data mining, natural language processing, and machine learning, along with learning analytics, to offer a wide range of features, including five components: (1) Student academic performance and behavior prediction and intervention, (2) Early warning system (EWS), (3) Analytics of Individualized Education Plan (IEP), (4) Talented students' prediction and identification, and (5) Cross-school personalized electives recommendation. We explore the following questions: (1) Is it feasible to implement a comprehensive and cohesive system that effectively uses AI to support teaching and learning in a diverse range of primary and secondary schools? (2) What about the user experience of the system in the real world? (3) What are the challenges of using this kind of system in the real world? (4) How can the system be promoted and what is its sustainability?

The development of this AI_aided K-12 system has been meticulously carried out while prioritizing user privacy and addressing the challenges posed by data heterogeneity. We successfully implemented the DMP_AI system in real-world primary and secondary schools. The user experience survey shows a generally positive response from the participants. Users perceive that the system is helpful in real-world educational scenarios. However, there are also some challenges to using this kind of system in the real world. Specially since the level of understanding of AI technology varies from user to user, how to train, explain and support them to understand AI technology and the AI modules in our system and use AI effectively to assist teaching and learning is a challenge and a long-term task that needs to be carried out continuously.

This work allows us to gain valuable insight into the potential and challenges of integrating AI into K-12 education in the real world. This system will serve as a valuable resource for supporting educators in providing effective and inclusive K-12 education.

## 2 Related Work

Over the last five years, AI-aided systems and applications have begun to show promise in K-12 education, aiming to enhance the learning experience and outcomes for students [6]. These systems take advantage of the power of AI to provide personalized learning experiences and additional resources for students and teachers, automate administrative tasks, and offer predictive analytics to guide educational strategies.

### 2.1 Predictive Analysis

Academic performance prediction and intervention, as well as predicting and intervening in academic and learning behavior, can benefit from the use of machine learning (ML) and data analysis techniques. These approaches enable educators to make informed decisions and provide personalized support for improved educational outcomes. Harvey and Kumar (2019) [7], Cruz-Jesus et al. (2020) [8], and Costa-Mendes et al. (2021) [9] demonstrate the use of ML techniques such as Naive Bayes, Random Forest, and Support Vector Regression to predict SAT Math scores and understand students' academic performance. Tuba and Pelin (2022) employ a Beta Regression model to predict students' success rates in entrance examinations [10]. For the prediction and intervention of behaviors, Sansone (2019) [11] and Mnyawami et al. (2022) [12], use supervised ML to predict student dropouts. Zhuang and Gan (2017) [13] employ an ensemble ML model to forecast enrollment numbers for effective resource planning. Li et al. (2020) [14] help identify at-risk students in online learning environments using a customized ML approach. Rasheed et al. (2021) [15] propose a supervised ML classifier to identify at-risk children and design appropriate interventions. Ni et al. (2020) [16] develop natural language processing and ML technologies to automate risk assessment for school violence.

These studies highlight how ML and data analysis techniques can enable educators to predict academic performance, identify at-risk students, and intervene early with targeted support, resulting in improved educational outcomes.

### 2.2 Early Warning System (EWS)

Integrating AI with intelligent tutoring systems has also shown promise in the development of Early Warning Systems. According to a 2016 report based on a survey by the US Department of Education, slightly more than half of public high schools in the United States had implemented these systems [17]. The use of early warning systems is even more prevalent in higher education, with an estimated 90 percent of four-year institutions having some form of system in place, as reported in a study conducted in 2014 [18]. However, in the earlier stages, such as primary school, EWS are less common.

Chung and Lee (2019) developed an EWS using supervised ML to identify high school students at risk of dropping out in advance and provide them with necessary support [19]. Ben et al. (2023) proposed an algorithm that automatically generates early and accurate alerts for teachers of at-risk of failure learners [20]. The algorithm uses both an original concept of the alert rule to define the alerting method and temporal evaluation metrics to identify the reliable starting time for generating alerts. As a proof of concept, they apply the algorithm to four different EWSs using real data from k-12 learners enrolled in online learning courses. However, it is merely an algorithm and does not implement an actual early warning system. Amal et al. (2022) proposed a new alert algorithm of an educational EWS for generating risk alerts at the earliest [21]. The algorithm is based on a weekly prediction model that aims to generate early alerts. They used data from k-12 learners enrolled in an online physics-chemistry module. Similarly, they did not implement an EWS system.

### 2.3 Learning Analytics

Increased adoption of educational technologies, the emergence of concepts in the digital classroom, and the interest in big data innovations have led to a growing awareness of the potential implementation of learning analytics to support learning development in educational institutions. However, most studies have focused on the implementation of learning analytics in higher education [22-24]. Therefore, there is a need to understand better the implementation of learning analytics in primary and secondary [25].

In addition to analyzing the learning of general students, it is also crucial to conduct learning analytics for students with the Individualized Education Plan (IEP). As the proportion of IEP students continues to increase (the number of SEN pupils in Hong Kong local schools reached 58,890 in 2021-22, 11 percent of the student population [26]), ensuring they receive appropriate support and personalized education becomes paramount. This is necessary to ensure that they receive the necessary personalized support and education to help them reach their full potential and achieve academic and socioemotional success.

### 2.4 Talented/Gifted Students Prediction and Identification

Generally gifted and talented children have exceptional achievement or potential in one or more of the following domains [27]: (1) a high level of measured intelligence; (2) specific academic aptitude in a subject area; (3) creative thinking; (4) superior talent in visual and performing arts; (5) natural leadership of peers; and (6) psychomotor ability - outstanding performance or ingenuity in athletics, mechanical skills or other areas requiring gross or fine motor coordination. Usually, gifted and talented children are those identified by professionally qualified persons who by virtue of outstanding abilities, are capable of high performance.

Gifted education is a field where ML has yet to be utilized, even though one of the underlying problems of gifted education is classification, which is an area where learning algorithms have become exceptionally accurate [28].

### 2.5 Recommender System

AI can recommend learning resources, interventions, and course selection based on a student's interests, performance, and learning goals. However, in the real world, privacy concerns often hinder data sharing, as sensitive information about students and their academic records must be protected. Existing methods, such as collaborative filtering [29], content-based filtering [30], and matrix factorization [31], often struggle to model the sparse data and diverse information available within individual schools. The limited data available for each school can cause problems with overfitting and cold start, decreasing the quality of recommendations [32]. In addition, these approaches fail to capture the intricate relationships and diverse patterns that are prevalent in cross-school scenarios, ultimately leading to suboptimal recommendations.

In summary, previous research has begun to explore the potential of different AI applications in K-12 education. Using machine learning, natural language processing, educational data mining, and intelligent tutoring systems, researchers have made initial progress in developing systems that can benefit students and educators. However, more research, including more extensive studies, is needed to refine these methods and realize their true impact on K-12 education. Additionally, it is important to ensure effective implementation in real-world educational settings.

## 3 Methodology

To bridge these gaps and explore the full potential of different AI applications in K-12 education, we have designed and implemented the DMP_AI (Data Management Platform_Artificial Intelligence) system, an innovative AI-aided educational system specifically designed for K-12 education.

Our system has a diverse range of data sources, including numerical and textual information, sourced from targeted K-12 schools. This comprehensive dataset encompasses various aspects, such as students' demographic details, academic records (including in-school performance and public examination data), behavioral information (attendance, punishments, awards, homework submissions), extracurricular activities, Individualized Education Plan (IEP) data, and other relevant teaching and learning information. To ensure data security and privacy, we employ a trusted data storage system. Sensitive data are stored within a school-based storage infrastructure, where it undergoes de-identification and encryption processes. Subsequently, some data is centrally stored in a secure cloud platform.

Our DMP_AI utilizes data mining, natural language processing, and machine learning (such as federated learning, attention mechanism, and personalized recommendations), along with learning analytics. It goes beyond by integrating the five functionalities mentioned above and addressing the associated challenges in real-world scenarios. The system comprises five key components: (1) Student academic performance and behavior prediction and intervention, (2) Early warning system, (3) Analytics of Individualized Education Plan, (4) Talented students' prediction and identification, and (5) Cross-school personalized electives recommendation. in addition, the part of the system that focuses on IEP intervention recommendations is currently under development.

What's more, the system actively involves teachers in its development and implementation process. It is an iterative system, where multiple users actively engage and contribute to its enhancement. Upon utilizing the modules, users are encouraged to provide evaluations and feedback, which serve as valuable input for system refinement. Through this iterative process, the system continuously evolves and improves based on user input, creating a dynamic feedback loop that drives ongoing enhancements. The system integrates multi-modal data from various sources and provides a secure environment with Federated Learning for educators, students, and parents to track performance and receive cross-school personalized recommendations. Taking advantage of these capabilities, our goal is to overcome the obstacles in implementing AI in K-12 education and maximize the benefits for all involved stakeholders. The structure of DMP_AI is illustrated in Fig. 1.

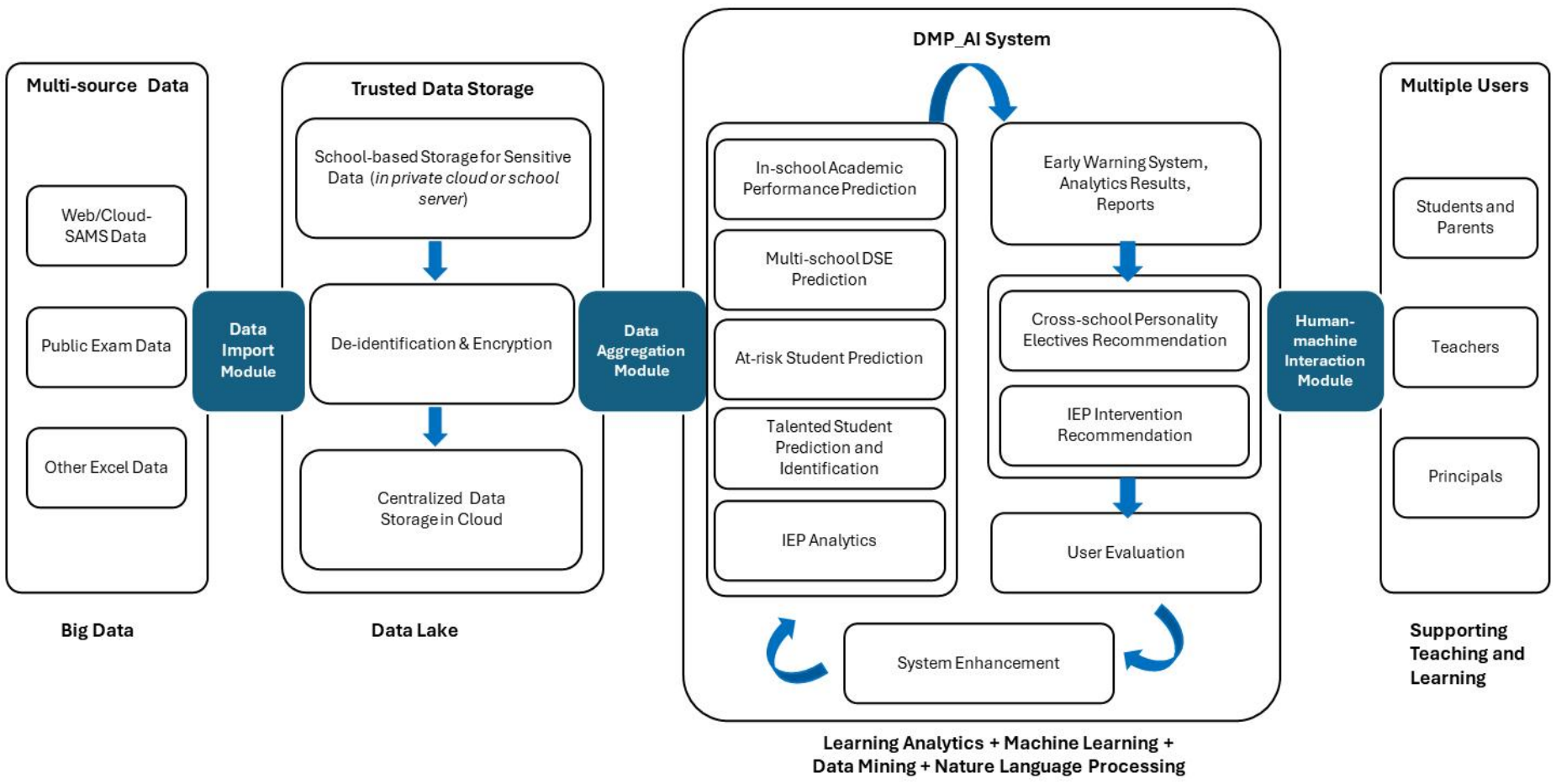

**Fig. 1.** The structure of DMP_AI. Web/Cloud-SAMS: Web/Cloud-based School Administration and Management System, Education Bureau, Hong Kong.

### 3.1 Student Academic Performance and Behavior Prediction and Intervention

Our system aims to provide a comprehensive view of each student's progress by offering an integrated solution for student academic performance and behavior prediction and intervention.

**Student Academic Performance Prediction and Intervention** Using digital technologies to track and predict student learning outcomes has emerged as a prominent trend in the field of AI applications in education. Our system incorporates student in-school academic performance prediction (Fig. 2 (a)), and public examination score prediction (Fig. 2 (b)) to fulfill different requirements.

A fundamental aspect of our system is its ability to predict student academic performance in the school. Through the integration of AI algorithms and data analysis techniques, we analyze a wide range of factors, including past academic records, attendance, punishments and awards, homework submission records, extracurricular activities, and other relevant demographics. Using these inputs, our system can accurately predict students' future academic performance in different subjects. These predictions are then presented to educators, providing them with valuable insights into each student's progress, understanding their current situation, and identifying potential areas for improvement. In addition to in-school academic performance prediction, our system also focuses on predicting students' performance in public examinations, such as the Hong Kong Diploma of Secondary Education (DSE) examination. These performances play a crucial role in assessing students' overall academic proficiency and often have significant implications for their future educational opportunities. By using historical data, student learning profiles, and performance patterns, our AI-powered system can predict students' upcoming scores in multisubject public examinations. The predicted scores/grades are presented through the user interface using different colors, allowing educators to identify students who may need additional support and customized interventions starting from the fourth year of secondary school, which is earlier than other existing systems, fostering a proactive approach to academic success.

**Student Behaviors Prediction and Intervention** We understand that student success goes beyond academic performance and encompasses their overall behavior and engagement in the learning process.

Through advanced AI algorithms and data analytics techniques, we not only track and predict students' academic achievements, but also monitor and analyze their behavior beyond academic performance (Fig. 2(c)). Taking into account factors such as attendance, participation in extracurricular activities, disciplinary records, and other relevant behavioral indicators, we provide a comprehensive predictive view of each student's behavioral progress and trends. These predictive results enable educators to identify students who may need additional support or intervention. With this information, educators can refer to various student information within the system to customize intervention measures according to specific needs, creating a more personalized learning experience for each student.

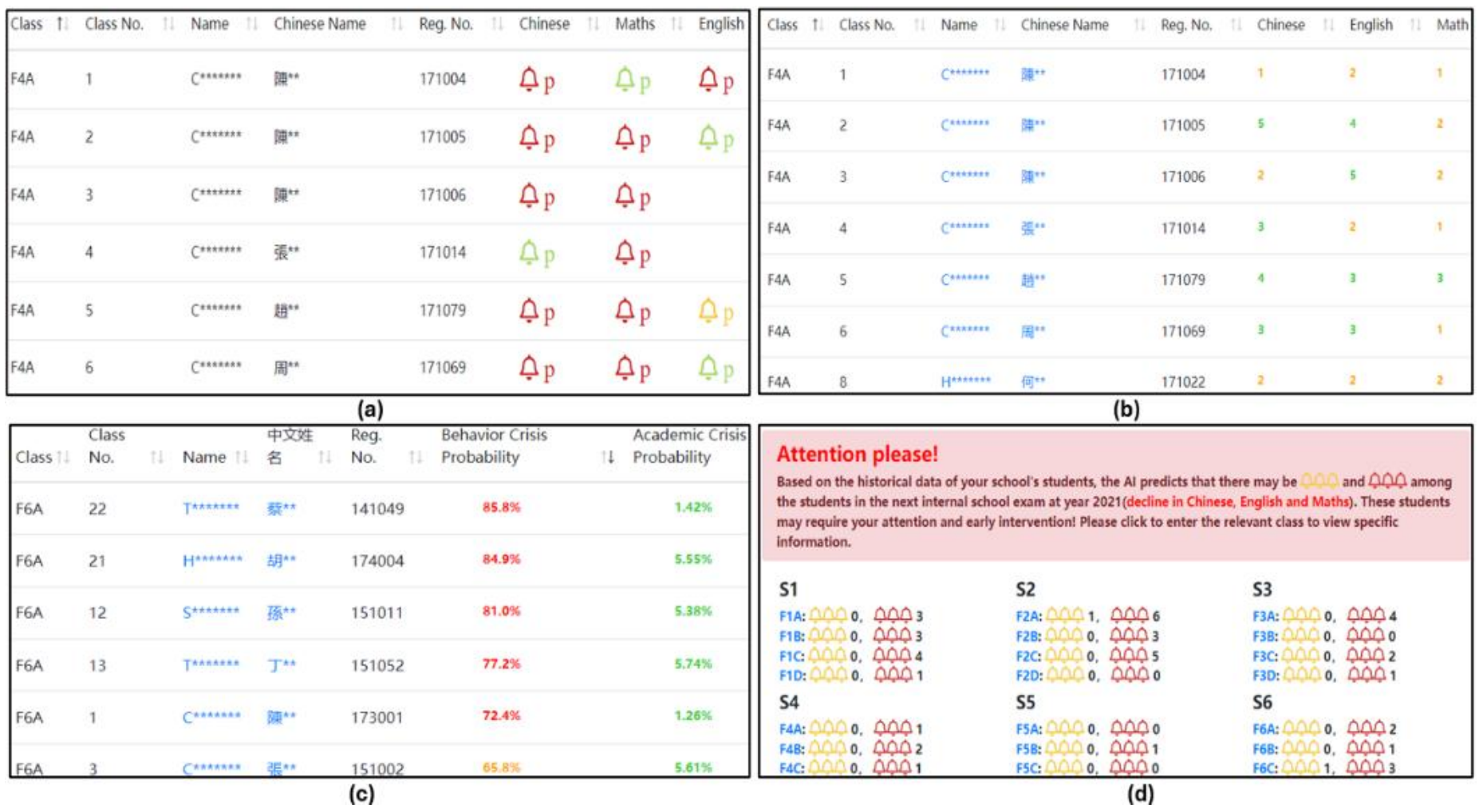

**Fig. 2.** Predictive Analysis and EWS. (a) In-school academic performance prediction and intervention; (b) Public examination performance prediction and intervention; (c) & (d) Behaviors prediction & EWS.

### 3.2 Early Warning System

Based on the predictive results mentioned above, our system has expanded an EWS to incorporate teacher feedback by involving them in the development process, proving display and alert mechanisms for analysis and prediction results. Firstly, we have added more indicators and graphs to the interface to provide a comprehensive and detailed view of student performance and trends. In the EWS (Fig. 2(c), (d)), we have included other key indicators such as homework completion, attendance, disciplinary actions, and more. The analysis and presentation of these indicators will help teachers gain a better understanding of students' overall learning status and behavioral characteristics. Secondly, we have further refined the alert mechanisms to accurately reflect student performance and trend changes through different levels and degrees. For in-school academic performance predictions, we have set different levels of indicators to more precisely differentiate the extent of score changes. For example, a red indicator indicates a significant decline in scores, a yellow indicator represents a slight decline, and a green indicator suggests that scores will remain stable or improve. In the prediction of public examination scores, green indicates that the student's expected score meets the requirements, yellow indicates a slight deviation, and red indicates a significant deviation. In terms of behavior prediction, a red indicator signifies a significant crisis in student behavior that requires immediate action, a yellow indicator suggests a potential crisis that requires the teacher's attention, and ongoing observation, while a green indicator indicates normal behavior with no crisis. Furthermore, we provide personalized settings options that allow teachers to adjust the thresholds of red and green indicators according to their specific needs. This flexibility enables customization to align with different school standards and subject requirements.

Through these enhancements and extensions, our system offers a more comprehensive and precise alert system, helping teachers gain a better understanding of student academic performance and behaviors. This, in turn, enables them to take timely intervention measures, providing targeted support and guidance to improve students' academic achievements and overall development.

### 3.3 Analytics of Individualized Education Plan (IEP)

With the increasing number of students with special educational needs, more and more research is focusing on how to customize individual education plans for students with different special educational needs. Our system uses cross-school student IEP data to conduct in-depth analysis, comparison, and presentation of the commonalities and characteristics of students with different types of special educational needs in terms of their learning performance, needs, and preferences. This provides assistant tools for IEP counselors to develop comprehensive and effective personalized education plans. For example, the system extracts part-of-speech tagging of text information related to students' special educational needs and learning situations. It filters out phrases that are deemed uninformative based on predefined phrase part-of-speech rules and extracts key information to generate word clouds. This provides an overview of the special educational needs of students and learning situations of students for the teachers (Fig.3(a), (b)). Through ML, data mining, heatmaps and correlation analysis, the system uncovers the correlation between extracurricular activity categories and different types of special educational needs students participate in. This facilitates teachers in designing personalized learning plans for individual students. What's more, we are actively researching a cross-school personality intervention recommender system that utilizes IEP analytics. This groundbreaking initiative aims to provide personalized intervention recommendations for students in different schools, taking into account their unique personalities and educational needs.

### 3.4 Talented Students' Prediction and Identification

Identifying talented and gifted students is a crucial aspect of K-12 education. These students have exceptional abilities and require specialized educational interventions to fully develop their talents. Traditionally, identification processes have relied on subjective assessments and standardized tests, which may overlook certain indicators of giftedness. However, advancements in artificial intelligence (AI) technologies present new opportunities for more effective and comprehensive identification methods. We use AI algorithms and data analytics techniques to conduct a comprehensive analysis of various factors that contribute to student talent identification. These factors include past academic records, attendance, awards, extracurricular activities, and relevant demographic information. Our system takes a data-driven approach to reveal students' capabilities across different academic and skill domains, presenting a detailed and extensive list of talented students in seven distinct categories: academic, sports, arts, leadership, service, technology, and others (Fig. 3(c)). By providing this information in a structured and organized manner, our system offers valuable support to educators in catering to

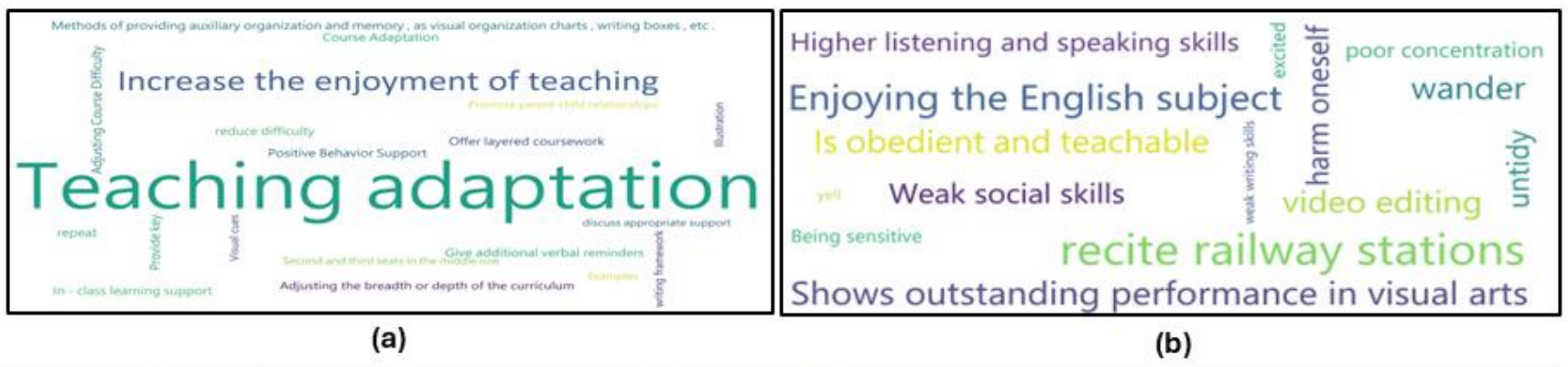

| Class | Class No. | English Name | Chinese Name | Reg. No. | Chinese rank | Maths rank | English rank | Academics | Sports | Arts | Leadership |
|---|---|---|---|---|---|---|---|---|---|---|---|
| F6A | 1 | C******* | 陳** | 173001 | 4 | 25 | 14 | 5 | 0 | 0 | 5 |
| F6A | 2 | C******* | 鄺** | 151001 | 2 | 65 | 33 | 18 | 0 | 1 | 6 |
| F6A | 3 | C******* | 張** | 151002 | 25 | 76 | 43 | 8 | 5 | 0 | 2 |
| F6A | 5 | F******* | 方** | 151004 | 82 | 53 | 55 | 21 | 0 | 1 | 10 |
| F6A | 6 | G******* | 吳** | 151005 | 96 | 41 | 82 | 12 | 0 | 1 | 8 |

(c)

**Fig. 3.** IEP analytics, Talented students' prediction and identification. (a) Overview of special education needs; (b) Overview of student learning situations; (c) Talented students' prediction and identification.

the diverse needs of students and nurturing their individual potential. The insights derived from our data-driven approach empower educators to better understand and address the unique talents of their students, ensuring a more personalized and effective educational experience.

### 3.5 Cross-school Personalized Electives Recommendation

Addressing cross-school learner diversity is crucial, especially in personalized recommendation systems for elective course selection. However, privacy concerns often limit cross-school data sharing, which hinders existing methods' ability to model sparse data and address heterogeneity effectively, ultimately leading to sub-optimal recommendations. In response, we propose HFRec [33], a heterogeneity-aware hybrid federated recommendation system designed to recommend cross-school electives. The proposed model constructs heterogeneous graphs for each school, incorporating various interactions and historical behaviors between students to integrate context and content information. We design an attention mechanism to capture heterogeneity-aware representations. Moreover, under a federated scheme, we train individual school-based models with adaptive learning settings to recommend tailored electives (Fig. 4).

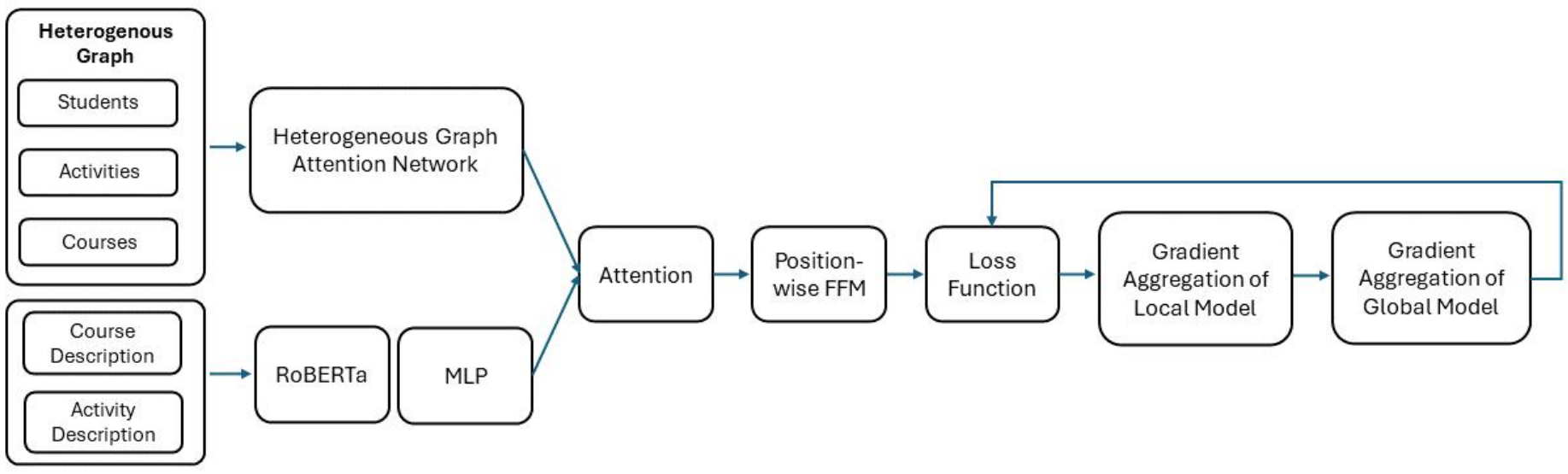

**Fig. 4.** The framework of HFRec.

## 4 Results

After the completion of development, starting from March 2023, four AI modules within the DMP_AI system have been installed and piloted in eight local schools, including three primary schools and five secondary schools. During the pilot phase, we conducted a survey among participants from diverse educational backgrounds to evaluate the real-world effectiveness and user satisfaction of our AI system. The survey consisted of ten questions that covered various aspects such as overall performance, user interface, system operation, and others. Participants were asked to rate each question on a scale of 1 to 5, with 1 being the lowest rating and 5 being the highest rating. The overall score is the average score for these four modules. In this survey, the M1 and M2 (Table 1) modules cover both version 1 and version 2, while the M3 and M4 (Table 1) modules only cover version 1.

**Table 1.** User experience. Rate (1-5). M1: In-school performance prediction. M2: Public examination prediction. M3: EWS. M4: Talented students' prediction and identification.

| Questions | M1 | M2 | M3 | M4 | Overall |
|---|---|---|---|---|---|
| I am satisfied with the overall performance. | 3.65 | 2.25 | 3.71 | 2.50 | 3.03 |
| I am satisfied with the user interface experience. | 3.75 | 3.25 | 4.43 | 4.00 | 3.86 |
| I think I have mastered how to use the module. | 3.55 | 3.50 | 3.71 | 4.00 | 3.69 |
| I believe that the installed module is helpful. | 3.80 | 3.00 | 3.86 | 3.50 | 3.54 |
| I have gained a better understanding of AI. | 3.20 | 3.00 | 2.86 | 2.50 | 2.89 |
| I can use it with the user manual, without training. | 3.05 | 3.75 | 3.29 | 2.00 | 3.02 |
| I would like to use the module at home. | 3.45 | 3.25 | 3.00 | 3.50 | 3.30 |
| I will continue to use the module in the future. | 3.70 | 3.25 | 3.57 | 3.50 | 3.51 |
| I will recommend this module to others. | 3.75 | 3.00 | 3.71 | 3.00 | 3.37 |
| I want to explore new AI modules. | 3.90 | 3.50 | 4.00 | 2.50 | 3.48 |

Table 1 gives a summary of feedback from 33 users about their experience using DMP_AI. In general, the results indicate a generally positive response from the participants. The average rating across nine questions is above 3.0, suggesting a satisfactory level of user experience and perception of the installed system. The participants expressed the highest satisfaction with the user interface experience, with an average rating of 3.86. This indicates that the design and functionality of the system's interface were well-received by the users. In terms of overall performance, the average rating was 3.03, indicating a moderate level of satisfaction. Although there is room for improvement, the feedback suggests that the system performance was deemed acceptable by the participants. The perceived helpfulness of the installed module garnered a commendable rating of 3.54, underscoring the users' recognition of its considerable benefits. Participants also expressed interest in further exploring AI modules, with an average rating of 3.48 for their interest in more newly developed AI modules. This indicates a positive inclination towards expanding their knowledge and engaging with AI technologies. However, the overall score for the question 'I have gained a better understanding of AI' is 2.89, below 3.0. This result indicates that participants have a relatively lower level of understanding of AI through the module. It may be necessary to further explore the reasons behind this and consider improving training or providing more detailed information to enhance participants' understanding of AI. The table shows that several questions related to the prediction and identification of talented students received lower ratings compared to the other three modules. The reasons for this could be related to the following factors: Users are trying to use AI technology for the first time to replace traditional identification processes. Users lack experience in utilizing AI technology to assist in identifying talented students. Users may have lacked a clear definition of talented students. The interdisciplinary nature of talent means that talent can manifest in various domains, including but not limited to sports, arts, technology, and leadership. Limited availability of data: the data is incomplete or inconsistent across different schools. Data diversity: the data collected from different schools exhibit heterogeneity, making it difficult to develop a unified model that can accurately capture the complexity and diversity of the data. Therefore, developing a model that can effectively predict talent across different fields and gain general understanding and acceptance among users is a challenging task.

In summary, the survey results reflect a generally positive response from the participants, highlighting their satisfaction with the user interface experience and their interest in AI modules. Users perceive that the system is helpful in real-world educational scenarios. They express a willingness to continue using the module in the future and recommend it to others. The feedback collected will provide invaluable insights that will not only enhance the current system but also contribute to the development of future AI modules. However, understanding AI among users from different backgrounds is a challenge, and further training and education are needed to enhance their understanding and knowledge.

## 5 Conclusion

To address the gaps in AI application in K-12 education, we have designed and implemented the DMP_AI system, an innovative AI-aided educational system that uses data mining, natural language processing, and machine learning, along with learning analytics, to offer a wide range of features for K-12 education in the real world. The development of this system has been meticulously carried out while prioritizing user privacy and addressing the challenges posed by data heterogeneity. To ensure the effectiveness of our system, we actively involved teachers as part of our development team. Their expertise and insight have greatly contributed to the modules' design and implementation. We have piloted four AI modules in eight local schools, and the feedback from the participants indicates a satisfactory level of satisfaction with the overall performance, user interface experience, and usability of the installed module. Participants have recognized the effectiveness of the system in real-world educational scenarios. They also expressed a strong interest in further exploring more AI modules and interest in using the module at home. However, some areas require attention and improvement. Participants have indicated a relatively lower rating when it comes to gaining a better understanding of AI through the modules. Regarding technical support and training needs, users expressed a moderate level of self-sufficiency, with a rating of 3.02. This highlights the need to evaluate the effectiveness of the modules in delivering comprehensive AI knowledge. This indicates additional technical support, resources, and training to enhance participants' understanding, although the cost of such training may pose a challenge in the future.

The findings highlight the importance of addressing user concerns, improving the training content, and providing additional technical support and training to enhance the user experience and ensure the long-term success and sustainability of the AI system. In the future, more innovative AI modules will be integrated into our DMP_AI and deployed to local schools, for example, the cross-school IEP intervention recommendation. In addition to the desktop version, the DMP_AI system will provide the mobile and tablet versions for convenient user access. At the same time, it is essential to integrate ethics into our AI development process. As we move forward, we will actively involve ethical considerations to ensure responsible and ethical AI practices within our system. This will help address potential biases, fairness, and transparency issues associated with AI algorithms and data processing. Addressing these areas of concern will contribute to enhancing AI applications in K-12 education effectiveness and ensuring a more satisfactory user experience.

**Acknowledgments.** This work is supported by Hong Kong Jockey Club Charities Trust (Project S/N Ref.: 2021-0369), and the Research Institute for Artificial Intelligence of Things, The Hong Kong Polytechnic University.